\documentstyle[a4,12pt,epsf]{article}

\newcommand{\be}{\begin{equation}}
\newcommand{\ee}{\end{equation}}
\newcommand{\epg}{$\eta \rightarrow \pi^{0} \gamma \gamma \;$}
\newcommand{\sla}{\hspace{-0.5 em}/}

\begin{document}
\begin{flushright}
TIT/HEP-318/NP \\
INS-Rep.-1132 \\
\end{flushright}
\begin{center}
{\huge \epg decay in the three-flavor Nambu--Jona-Lasinio model}\\ 
\vspace{2em}
{\large Y.Nemoto
\footnote{E-mail address: nemoto@th.phys.titech.ac.jp}
 , M.Oka
\footnote{E-mail address: oka@th.phys.titech.ac.jp}
} \\ 
  {\large \em Department of Physics, Tokyo Institute of Technology,} \\
  {\large \em Meguro, Tokyo 152, Japan} \\ 
\vspace{1em}
{\large M.Takizawa}
\footnote{E-mail address: takizawa@ins.u-tokyo.ac.jp}
 \\ 
  {\large \em Institute for Nuclear Study, University of Tokyo,} \\
  {\large \em Tanashi, Tokyo 188, Japan}
\end{center}
\vspace{2em}

\baselineskip 2em
\begin{abstract}
\baselineskip 2em
We study the \epg decay via the quark-box diagram in the three-flavor 
Nambu--Jona-Lasinio model that includes the $U_{A}(1)$ breaking effect. 
We find that the $\eta$-meson mass, the $\eta \to \gamma \gamma$ decay
width and the \epg decay width are in good agreement 
with the experimental values when the $U_{A}(1)$ breaking is strong
and the flavor $SU(3)$ singlet-octet mixing angle $\theta$ is 
about zero. The photon energy and the photon invariant 
mass spectra in \epg are compared with those in the chiral 
perturbation theory.
\end{abstract}
\clearpage

Chiral symmetry plays an essential role in the light-flavor QCD.
Light pseudoscalar mesons are regarded as the Nambu-Goldstone (NG)
bosons
of spontaneous symmetry breaking of chiral $SU(N)_L\times SU(N)_R$
symmetry. Properties of the NG bosons have been studied in various
chiral effective theories successfully.

Another interesting low-energy symmetry is the axial $U(1)$ symmetry,
which is explicitly broken by the anomaly.  The symmetry
breaking is manifested in the heavy mass of $\eta'$ meson, which has
been studied as the ``$U_A(1)$ problem''.  It is yet not clear how
strong the anomaly effect is on the pseudoscalar spectrum mainly
because of complicated interference of the explicit flavor symmetry
breaking due to the strange quark mass.  Two of the authors (M.T.\ and 
M.O.) argued in a previous paper\cite{Taki}\ that a strong
$U_A(1)$ breaking and consequently a large flavor mixing are favorable 
for the $\eta$ meson.  The argument is based on the analysis of the
$\eta$ system as a $q\bar q$ bound state in the three-flavor 
Nambu--Jona-Lasinio (NJL) model.
We claim that the $\eta$ mass as well as the $\eta\to
\gamma\gamma$ decay rate supports a $U_A(1)$ breaking six-quark 
interaction much stronger than previously used\cite{Hatsuda}.
This interaction causes a large flavor mixing resulting almost pure 
octet $\eta \simeq \eta_{8}$.  

The purpose of this paper is to
extend the previous analysis to another $\eta$ decay process,
$\eta\to\pi^0\gamma\gamma$ and to confirm our claim of the strong $U_A(1)$
breaking. We are interested in this process where the internal
structure of the $\eta,\pi^0$ mesons plays essential roles because the 
photon does not couple 
to the neutral mesons directly. Furthermore, chiral perturbation
theory (ChPT) gives too small prediction in the leading order and higher
order terms are expected to be dominant.

The three-flavor NJL model is one of the phenomenologically
successful chiral effective models of the low-energy QCD and the
lagrangian density we have used is as follows,
\begin{eqnarray}
  {\cal L} & = & {\cal L}_{0}+{\cal L}_{4}+{\cal L}_{6}, \\
  {\cal L}_{0} & = & \overline{\psi} (i \partial_{\mu} \gamma^{\mu} - \hat{m})
 \psi, \\
  {\cal L}_{4} & = & \frac{G_{S}}{2} \sum_{a=0}^{8} \left[ \left( 
\overline{\psi} \lambda^{a} \psi \right)^{2} + \left( \overline{\psi} 
\lambda^{a} i \gamma_{5} \psi \right)^{2} \right], \\
  {\cal L}_{6} & = & G_{D} \left\{ \mathop{\det}_{(i,j)} \left[ 
\overline{\psi}_{i} (1+\gamma_{5}) \psi_{j} \right] + \mathop{\det}_{(i,j)} 
\left[ \overline{\psi}_{i} (1-\gamma_{5}) \psi_{j} \right] \right\},
\end{eqnarray}
where $\psi$ is the quark field, $\hat{m}=diag(m_{u},m_{d},m_{s})$ is 
the current quark mass matrix, and $\lambda^{a}$ is the flavor $U(3)$ 
generator $(\lambda^{0}= \sqrt{2/3}I)$. The determinant in  ${\cal L}_{6}$ 
is a $3 \times 3$ determinant with respect to the flavor indices $i,j=u,d,s$. 
The model involves the $U_{L}(3) \times U_{R}(3)$ symmetric four-quark 
interaction ${\cal L}_{4}$ and the six-quark flavor-determinant interaction 
${\cal L}_{6}$ incorporating effects of the $U_{A}(1)$ anomaly. 

Quark condensates and constituent quark masses are self-consistently
determined by the gap equations in the mean field approximation.
The covariant cutoff $\Lambda$ is introduced to regularize the
divergent integrals. The pseudoscalar channel quark-antiquark
scattering amplitudes are then calculated in the ladder approximation.
From the pole positions of the scattering amplitudes, the pseudoscalar
meson masses are determined. We define the effective meson-quark 
coupling constants $g_{\eta q q}$ and $g_{\pi q q}$ by introducing 
additional vertex lagrangians,

\begin{table}[t]
\caption{The parameters and \epg decay widths for each $G_{D}^{\rm eff}$}
\begin{center}
\begin{tabular}{|c|c|c|c|c|c|c|} 
\hline
 $G_{D}^{\rm eff}$ & $G_{S}^{\rm eff}$ & $M_{\eta}$[MeV] 
& $\theta(M_{\eta}^{2})$[deg] & $g_{\eta qq}$ & $\Gamma$[eV] \\ \hline
0.00 & 0.73 & 138.1 & -54.74 & 3.44 & 2.88 \\
0.10 & 0.70 & 285.3 & -44.61 & 3.23 & 2.46 \\
0.20 & 0.66 & 366.1 & -33.52 & 3.12 & 2.06 \\
0.30 & 0.63 & 419.1 & -23.24 & 3.11 & 1.71 \\
0.40 & 0.60 & 455.0 & -14.98 & 3.15 & 1.42 \\
0.50 & 0.57 & 479.7 & -8.86  & 3.20 & 1.20 \\
0.60 & 0.54 & 497.3 & -4.44  & 3.25 & 1.04 \\
0.70 & 0.51 & 510.0 & -1.25  & 3.28 & 0.92 \\
0.80 & 0.47 & 519.6 &  1.09  & 3.30 & 0.84 \\
0.90 & 0.44 & 527.0 &  2.84  & 3.31 & 0.77 \\
1.00 & 0.41 & 532.8 &  4.17  & 3.32 & 0.71 \\
1.10 & 0.40 & 537.5 &  5.21  & 3.32 & 0.67 \\
1.20 & 0.35 & 541.3 &  6.02  & 3.31 & 0.63 \\
1.30 & 0.32 & 544.5 &  6.66  & 3.30 & 0.61 \\
1.40 & 0.29 & 547.2 &  7.17  & 3.29 & 0.58 \\
1.50 & 0.25 & 549.4 &  7.57  & 3.28 & 0.56 \\
1.60 & 0.22 & 551.4 &  7.90  & 3.26 & 0.55 \\
\hline
\end{tabular} 
\end{center}
\end{table}

\be
  {\cal L}_{\eta q q} = g_{\eta q q} \overline{\psi} i\gamma_{5}\lambda^{\eta}
\psi \phi_{\eta},
\ee
\be
  {\cal L}_{\pi  q q} = g_{\pi  q q} \overline{\psi} i\gamma_{5}\lambda^{3}
\psi \phi_{\pi^{0}}, 
\ee
with $ \lambda^{\eta}=\cos \theta \lambda^{8}-\sin \theta \lambda^{0}$.
Here $\phi$ is an auxiliary meson field introduced for convenience and 
the effective meson-quark coupling constants are calculated from the residues 
of the $q \bar{q}$-scattering amplitudes at the corresponding meson poles.
Because of the $SU(3)$ symmetry breaking, the flavor $\lambda^{8}-\lambda^{0}$
components mix with each other. Thus we solve the coupled-channel $q \bar{q}$
scattering problem for the $\eta$ meson. The mixing angle $\theta$ is 
obtained by diagonalization of the $q \bar{q}$-scattering amplitude
at the $\eta$-meson pole. The meson decay constant 
$f_{M} \; (M = \pi, K, \eta)$ is determined by calculating the quark-antiquark
one-loop graph. The explicit expressions are found in \cite{Taki}.

The parameters of the model are the current quark masses $m_{u}=m_{d},m_{s}$,
the four-quark coupling constant $G_{S}$, the six-quark determinant coupling
constant $G_{D}$ and the covariant cutoff $\Lambda$. We take $G_{D}$ as a 
free parameter and study $\eta$ meson properties as functions of $G_{D}$.
We use the light current quark masses $m_{u}=m_{d}=8.0$ MeV (same as in 
\cite{Taki}). Other parameters, $m_{s},\; G_{D}$, and $\Lambda$, are determined
so as to reproduce the isospin averaged observed masses, $m_{\pi}, m_{K}$,
and $f_{\pi}$.

We obtain $m_{s}=193$ MeV, $\Lambda=783$ MeV, the constituent $u,d$-quark
mass $M_{u,d}=325$ MeV and $g_{\pi q q}=3.44$, which are almost independent
of $G_{D}$.

Table 1 summarizes the fitted results of the model parameters and the 
quantities necessary for calculating the \epg decay width which depend on
$G_{D}$.  We define dimensionless parameters $G_{D}^{\rm eff} \equiv
- G_{D} (\Lambda / 2 \pi)^{4} \Lambda N_{c}^{2}$ and $G_{S}^{\rm eff} \equiv
G_{S} (\Lambda / 2 \pi)^{2} N_{c}$. When $G_{D}^{\rm eff}$ is zero, our
lagrangian does not cause the flavor mixing and therefore the ideal mixing
is achieved. The ``$\eta$'' is purely $u\bar u + d\bar d$ and is degenerate 
to the pion in this limit.
 
It is found in \cite{Taki} that the $\eta \rightarrow \gamma \gamma$ 
decay width is reproduced at about $G_{D}^{\rm eff}=0.7$. 
At this value the ratio $G_{D} \langle \overline{s} s \rangle / G_{S} =0.44$
indicates that the contribution from ${\cal L}_{6}$ to the dynamical mass of 
the up and down quarks is 44\% of that from ${\cal L}_{4}$.
The mixing angle at $G_{D}^{\rm eff}=0.7$ is $\theta = -1.3^{\circ}$ and 
that indicates a strong OZI violation and a large (u,d)-s mixing. 
This disagrees with the ``standard'' value $\theta \simeq -20^{\circ}$ 
obtained in ChPT\cite{Don}. This is due to the 
stronger $U_{A}(1)$ breaking in the present calculation.
The difference mainly comes from the fact that the mixing angle 
in the NJL model depends on $q^{2}$ of the $\overline{q}q$ state 
and thus reflects the internal structure of the $\eta$ meson. 
On the contrary the analyses of ChPT\cite{Don} 
assume an energy-independent mixing angle, i.e., 
$\theta(M_{\eta}^{2})=\theta(M_{\eta'}^{2})$.
 
We are now in a position to 
study whether  the \epg decay rate is consistent with our picture of 
$\eta$ with a large OZI mixing.

The experimental value of the \epg decay width is \cite{Exp}
\be
  \Gamma_{exp} \left( \eta \rightarrow \pi^{0} \gamma \gamma \right) 
= 0.85 \pm 0.19 {\rm \ eV}.
\ee
We evaluate the quark-box diagram given in Fig 1. 
We follow the evaluation of the box-diagram performed in \cite{Ng}.
Other possible contributions will be discussed later.

\input FEYNMAN
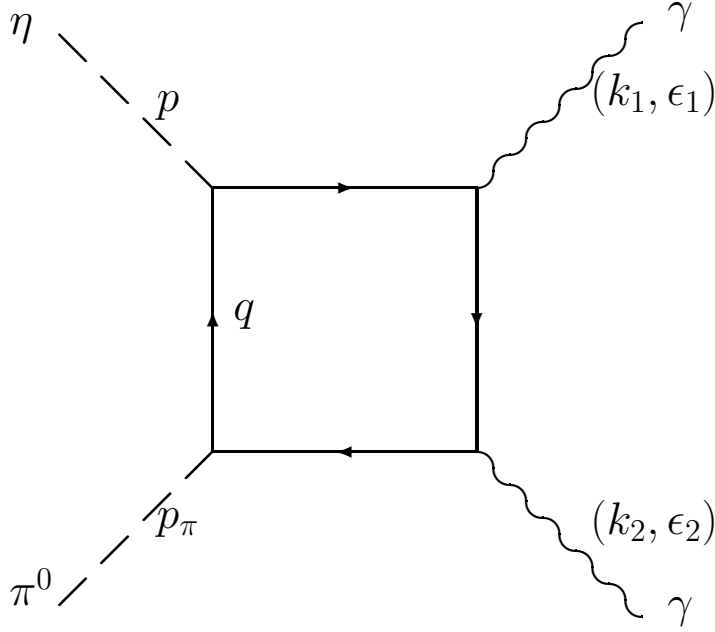
\begin{figure}[t]
\begin{picture}(40000,23000)

\thicklines\drawline\fermion[\N\REG](13000,8000)[10000]
\drawarrow[\N\ATBASE](\pmidx,\pmidy)
\put(\pmidx,\pmidy){\,\, \Large $q$}
\drawline\scalar[\NW\REG](\fermionbackx,\fermionbacky)[4]
\put(\pmidx ,\pmidy){\,\, \Large $p$}
\put(\pbackx,\pbacky){\hspace{-8 mm} \Large $\eta$}
\drawline\fermion[\E\REG](\fermionbackx,\fermionbacky)[\fermionlength]
\drawarrow[\E\ATBASE](\pmidx,\pmidy)
\drawline\photon[\NE\FLIPPED](\fermionbackx,\fermionbacky)[10]
\put(\pmidx,\pmidy){\,\,\,\, \Large $(k_{1},\epsilon_{1})$}
\put(\pbackx,\pbacky){\hspace{2 mm} \Large $\gamma$}
\drawline\fermion[\S\REG](\fermionbackx,\fermionbacky)[\fermionlength]
\drawarrow[\S\ATBASE](\pmidx,\pmidy)
\drawline\fermion[\W\REG](\fermionbackx,\fermionbacky)[\fermionlength]
\drawarrow[\W\ATBASE](\pmidx,\pmidy)
\drawline\photon[\SE\REG](\fermionfrontx,\fermionfronty)[10]
\put(\pmidx,\pmidy){\,\,\,\, \Large $(k_{2},\epsilon_{2})$}
\put(\pbackx,\pbacky){\hspace{2 mm} \Large $\gamma$}
\drawline\scalar[\SW\REG](\fermionbackx,\fermionbacky)[4]
\put(\pmidx,\pmidy){\,\, \Large $p_{\pi}$}
\put(\pbackx,\pbacky){\hspace{-8 mm} \Large $\pi^{0}$}

\end{picture}
\caption{The quark-box diagram for \epg}
\end{figure}

The \epg decay amplitude is given by
\be
  \langle \pi^{0}(p_{\pi}) \gamma(k_{1},\epsilon_{1}) \gamma(k_{2},
\epsilon_{2}) | \eta(p) \rangle = i (2\pi)^{4} \delta^{4}
(p_{\pi}+k_{1}+k_{2}-p) \epsilon_{1}^{\mu} \epsilon_{2}^{\nu} T_{\mu\nu},
\ee
where $\epsilon_{1}$ and $\epsilon_{2}$ are the polarization vectors of 
the photons. 
$T_{\mu\nu}$ is given by a straightforward evaluation of the Feynman diagrams. 
After calculating traces  in color and flavor spaces, we obtain
\be
  T_{\mu\nu}=-i \frac{1}{\sqrt{3}}(\cos \theta -\sqrt{2} \sin \theta )
e^{2}g_{\eta q q}g_{\pi q q} \int \frac{d^{4}q}{(2\pi)^{4}} 
\sum_{i=1}^{6} U_{\mu\nu}^{i},
\ee
with
\begin{eqnarray}
  U_{\mu\nu}^{1} & = & {\rm Tr}^{(D)} \left\{ \gamma_{5} \frac{1}{q \sla 
- M + i\epsilon} \gamma_{5} \frac{1}{q \sla + p \sla - k \sla_{1}  
- k \sla_{2} - M +i\epsilon} \right. \nonumber \\
  & \times & \left. \gamma_{\nu} \frac{1}{q \sla + p \sla - k \sla_{1} 
- M + i\epsilon} \gamma_{\mu} \frac{1}{q \sla + p \sla - M +i\epsilon} 
\right\} , \\
  U_{\mu\nu}^{2} & = & {\rm Tr}^{(D)} \left\{ \gamma_{5} \frac{1}{q \sla 
- M + i\epsilon} \gamma_{5} \frac{1}{q \sla + k \sla_{2} - M +i\epsilon} 
\right. \nonumber \\
  & \times & \left. \gamma_{\nu} \frac{1}{q \sla + p \sla - k \sla_{1} 
- M + i\epsilon} \gamma_{\mu} \frac{1}{q \sla + p \sla - M +i\epsilon} 
\right\} , \\
  U_{\mu\nu}^{3} & = & {\rm Tr}^{(D)} \left\{ \gamma_{5} \frac{1}{q \sla 
- M + i\epsilon} \gamma_{\nu} \frac{1}{q \sla + k \sla_{2} - M +i\epsilon} 
\right. \nonumber \\
  & \times & \left. \gamma_{\mu} \frac{1}{q \sla + k \sla_{1} + k \sla_{2} 
- M + i\epsilon} \gamma_{5} \frac{1}{q \sla + p \sla - M +i\epsilon} 
\right\} , \\
  U_{\mu\nu}^{4} & = & U_{\nu\mu}^{1}(k_{1} \leftrightarrow k_{2}), \\
  U_{\mu\nu}^{5} & = & U_{\nu\mu}^{2}(k_{1} \leftrightarrow k_{2}), \\
  U_{\mu\nu}^{6} & = & U_{\nu\mu}^{3}(k_{1} \leftrightarrow k_{2}).
\end{eqnarray}
Here  Tr$^{(D)}$ means trace in the Dirac indices and $M$ is the 
constituent u,d-quark mass. 
Because the loop integration in (9) is not divergent, we do not introduce
the UV cutoff. Then the gauge invariance is preserved. The inclusion of the
cutoff that is consistent with the gap equation will break the gauge 
invariance and make the present calculation too complicated.
Note that the strange quark does not contribute to the loop.

On the other hand the amplitude $T_{\mu\nu}$ has a general form required 
by the gauge invariance\cite{Eck}
\begin{eqnarray}
  T^{\mu\nu} & = & A(x_{1},x_{2})(k_{1}^{\nu} k_{2}^{\mu} - k_{1} \cdot k_{2} 
g^{\mu\nu}) \nonumber \\
  & + & B(x_{1},x_{2}) \left[ -M_{\eta}^{2} x_{1}x_{2} g^{\mu\nu} -
\frac{k_{1} \cdot k_{2}}{M_{\eta}^{2}} p^{\mu}p^{\nu} + x_{1} k_{2}^{\mu} 
p^{\nu} + x_{2} p^{\mu} k_{1}^{\nu} \right],
\end{eqnarray}
with
\be
  x_{i}=\frac{p \cdot k_{i}}{M_{\eta}^{2}},
\ee
and $M_{\eta}$ is the $\eta$ meson mass. With $A$ and $B$, 
the differential decay rate with respect to the energies of the two photons 
is given by
\begin{eqnarray}
  \frac{d^{2} \Gamma}{dx_{1} dx_{2}} & = & \frac{M_{\eta}^{5}}{256 \pi^{2}} 
\left\{ \left|  A  + \frac{1}{2} B \right|^{2} \left[ 2(x_{1}+x_{2})
+\frac{M_{\pi}^{2}}{M_{\eta}^{2}} -1 \right]^{2} \right. \nonumber \\
  &  + &  \left. \frac{1}{4} \left| B \right| ^{2} \left[ 4 x_{1} x_{2} 
- \left[ 2(x_{1}+x_{2})+ \frac{M_{\pi}^{2}}{M_{\eta}^{2}}-1 \right] 
\right] ^{2} \right\}  ,
\end{eqnarray}
where $M_{\pi}$ is the $\pi^{0}$ meson mass. 
Though the mass of $\eta$ as a $\bar{q} q$ bound state  depends on 
$G_{D}^{\rm eff}$,
we use the experimental value $M_{\eta}=547$ MeV in evaluating (18).
The Dalitz boundary is given by two conditions:
\be
  \frac{1}{2} \left( 1-\frac{M_{\pi}^{2}}{M_{\eta}^{2}} \right) \leq  x_{1}
+x_{2} \leq 1- \frac{M_{\pi}}{M_{\eta}} ,
\ee
and
\be
  x_{1}+x_{2}-2 x_{1} x_{2} \leq \frac{1}{2} 
\left(1-\frac{M_{\pi}^{2}}{M_{\eta}^{2}} \right) .
\ee
In evaluating  (10)-(15), one only has to identify the coefficients of 
$p^{\mu} p^{\nu}$ and $g^{\mu\nu}$. Details of the calculation are given 
in \cite{Ng}. Defining ${\cal A}$ and ${\cal B}$ by
\be
  \int \frac{d^{4}q}{(2\pi)^{4}} \sum_{i=1}^{6} U_{i}^{\mu\nu} 
= -i\left( {\cal A} g^{\mu\nu} + {\cal B} \frac{p^{\mu} p^{\nu}}{M_{\eta}^{2}}
 + \cdots \right) ,
\ee
we find $A$ and $B$ as
\begin{eqnarray}
  A & = & \frac{1}{\sqrt{3}} (\cos \theta - \sqrt{2} \sin \theta) e^{2} 
g_{\pi q q} g_{\eta q q} \frac{2}{M_{\eta}^{2} \sigma} 
\left[ {\cal A}- 2 x_{1} x_{2} \
\frac{{\cal B}}{\sigma} \right] , \\
  B & = & \frac{1}{\sqrt{3}} (\cos \theta - \sqrt{2} \sin \theta) e^{2} 
g_{\pi q q} g_{\eta q q} \frac{2}{M_{\eta}^{2}} \frac{{\cal B}}{\sigma} ,
\end{eqnarray}
with
\be
  \sigma = \frac{(k_{1}+k_{2})^{2}}{M_{\eta}^{2}} = 2(x_{1}+x_{2})
+\frac{M_{\pi}^{2}}{M_{\eta}^{2}} -1 .
\ee
We evaluate  ${\cal A}$ and ${\cal B}$ numerically and further integrate 
(18) to obtain the \epg decay rate. 
The results are given in the last column of Table 1 and shown in Fig 2.

\begin{figure}[t]
{\Large \mbox{\boldmath$\Gamma $} [eV]} \\
\epsfxsize=360 pt
\epsfbox{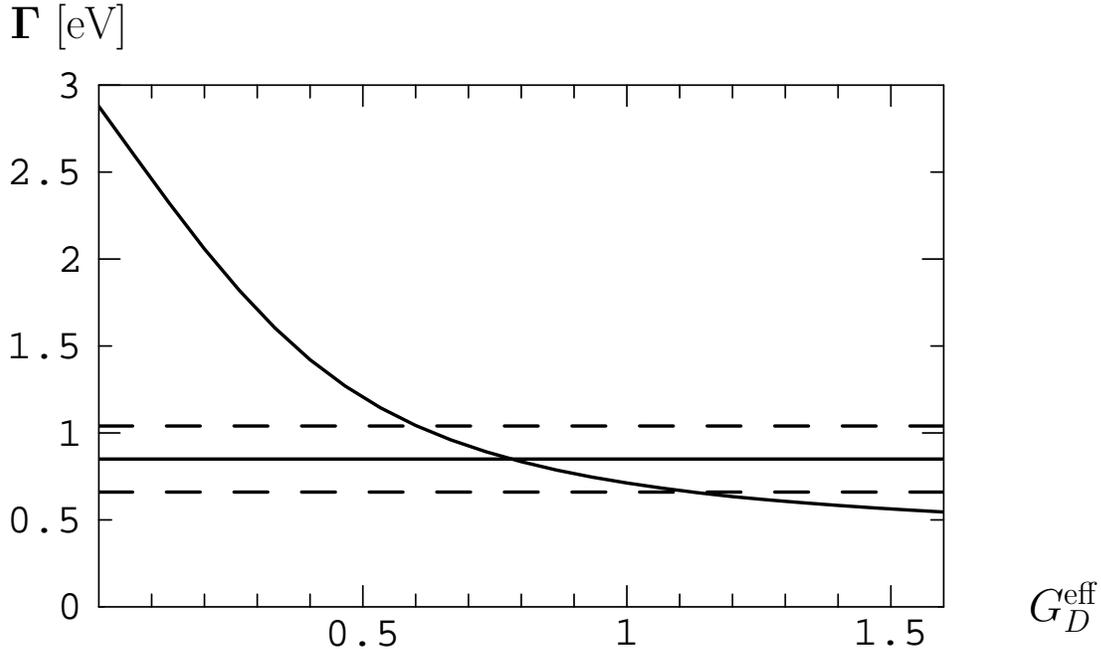} 
\begin{flushright}
  \vspace*{-15mm} 
  {\Large $G_{D}^{\rm eff}$}  
\end{flushright}
\caption{Dependence of the $\eta \to \pi^{0} \gamma \gamma $ decay
width on the dimension-less coupling constant $G_{D}^{\rm eff}$. The horizontal
solid line indicates the experimental value $\Gamma = 0.85$ eV and the dashed
lines indicate its error widths.}
\end{figure}

Our result is $\Gamma(\eta \rightarrow \pi^{0} \gamma \gamma )=0.92$ eV at 
$G_{D}^{\rm eff}=0.70$ where the  $\eta \rightarrow \gamma \gamma $ decay 
width 
is reproduced \cite{Taki}. At $G_{D}^{\rm eff}=1.40$ which reproduces 
the experimental $\eta$ meson mass,  
$\Gamma(\eta \rightarrow \pi^{0} \gamma \gamma )=0.58$ eV. 
Both are in reasonable agreement with (7).

In ChPT \cite{Ame}, there is no lowest order 
$O(p^{2})$ contribution to the \epg process because the involved mesons 
are neutral. Likewise the next order $O(p^{4})$ tree diagrams  do not exist. 
Thus the $O(p^{4})$ one-loop diagrams give the leading term in this process, 
but the contribution is two orders of magnitude smaller than the 
experimental value (7). This is because the pion loop violates the G-parity 
invariance and the kaon loop is also suppressed by the large kaon mass.

At $O(p^{6})$, there exists contribution coming from tree diagrams, 
one-loops and two-loops. The loop contributions are smaller than those 
from the order $O(p^{4})$. At $O(p^{8})$, more tree diagrams and a new type 
of loop corrections appear, but the loop corrections are also small.

In \cite{Ame}, coupling strengths of the tree diagrams are determined 
assuming saturation  by meson resonance poles, such as $\rho, \omega, a_{0}$ 
and $a_{2}$. This gives 
\be
  \Gamma(\eta \rightarrow \pi^{0} \gamma \gamma)=0.42 \pm 0.20 \;\; {\rm eV},
\ee
where the $0.20$ eV is the contribution of  $a_{0}$ and $a_{2}$, 
the sign of which is not known. The result is a factor two smaller 
than the experimental value.

The contributions of other mesons,  such as 
$b_{1}(1235),h_{1}(1170),h_{1}(1380)$ \cite{Ko2} 
and other tree diagrams \cite{Ko}, are found to be small.

Although these results based on ChPT are not too
far from the experimental value, it is noted that the higher order $O(p^{6})$
terms in the perturbation expansion are larger than the leading $O(p^{4})$
terms and the results contain ambiguous parameters that cannot be determined
well from other processes.

On the other hand in \cite{Bel} the $O(p^{6})$ tree diagrams are evaluated by 
using the extended NJL (ENJL) model\cite{Bijnens}. They calculated three
contributions in ENJL, namely, the vector and scalar resonance exchange and
the quark-loop contributions. Their result is $\Gamma(\eta \to \pi^0 \gamma
\gamma) \simeq 0.5$ eV. They further introduced the $O(p^{8})$ chiral 
corrections as well as the axialvector and tensor meson exchange contributions,
and finally obtained $\Gamma(\eta \to \pi^0 \gamma \gamma) = 0.58 \pm 0.3$ eV.

The difference between our approach and that in \cite{Bel} are as follows.
The ENJL model lagrangian has not only the scalar-pseudoscalar four quark 
interactions but also the vector-axialvector four quark interactions.
However, the $U_{A}(1)$ breaking is not explicitly included in their model
and therefore the $\eta-\eta'$ mixing is introduced by hand with the mixing
angle $\theta = -20^{\circ}$. We stress that the introduction of the $U_{A}(1)$
breaking interaction is important to understand the structure of the $\eta$
meson. 

There is another difference. The coupling constants of the chiral effective
meson lagrangian predicted in the ENJL model are parameters of the Green
function evaluated at zero momenta. On the other hand we evaluate the 
quantities at the pole position of the mesons.

\begin{figure}[t]
{\Large \mbox{\boldmath$\frac{d \Gamma}{d m^{2}_{\gamma\gamma}}$} }
($\times 10^{-6}$
[MeV$^{-1}]$) \\
\epsfxsize=380 pt
\epsfbox{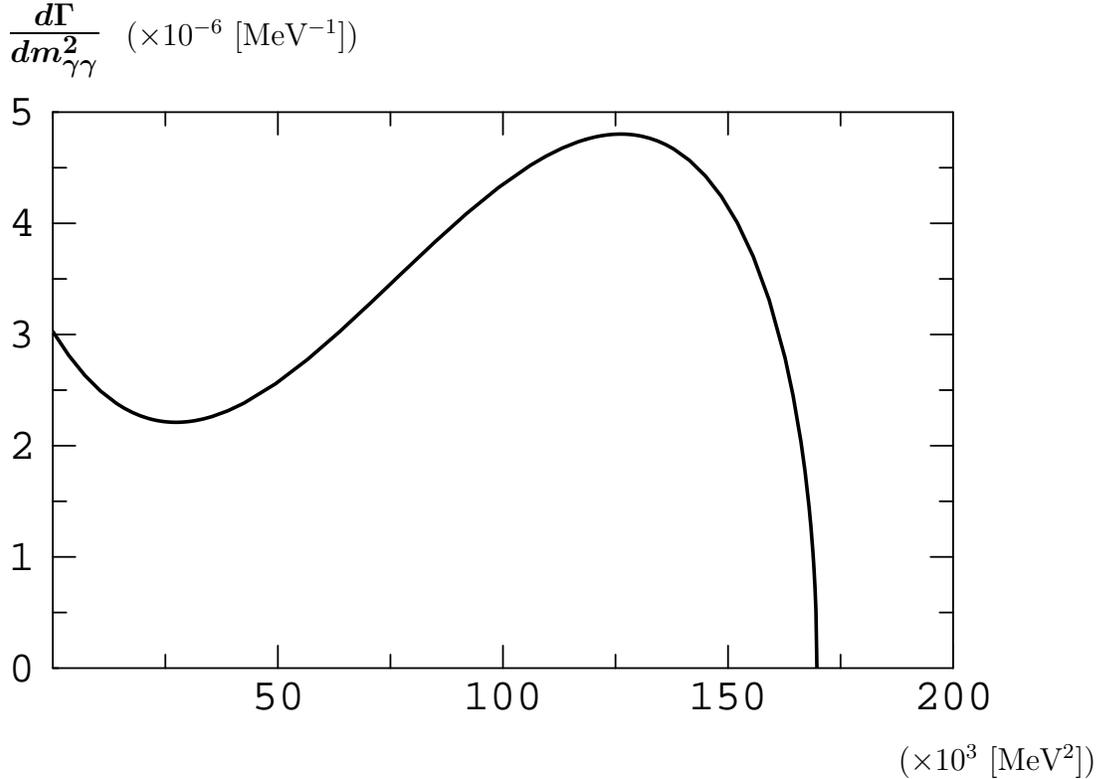}
\begin{flushright}
  \vspace*{-3mm}
  $(\times 10^{3}$ [MeV$^{2}])$
\end{flushright}
\caption{Spectrum of the photon invariant mass $m_{\gamma\gamma}^{2}$.}
\end{figure}

Calculated spectrum of the photon invariant mass square $m_{\gamma\gamma}^{2}$ 
for the \epg decay is shown in Fig 3. As this spectrum is compared with those 
calculated by ChPT in \cite{Ko}, 
we find ours to be similar to the one for 
$d_{3}=4.5 \times 10^{-2}$  GeV$^{-2}$ 
in \cite{Ko} which involves an additional $O(p^{6})$ contribution to 
the original Lagrangian. 

\begin{figure}[t]
{\Large  \hspace*{1cm} \mbox{\boldmath$\frac{d \Gamma}{d E_{\gamma}}$}} \\
\epsfxsize=380 pt
\epsfbox{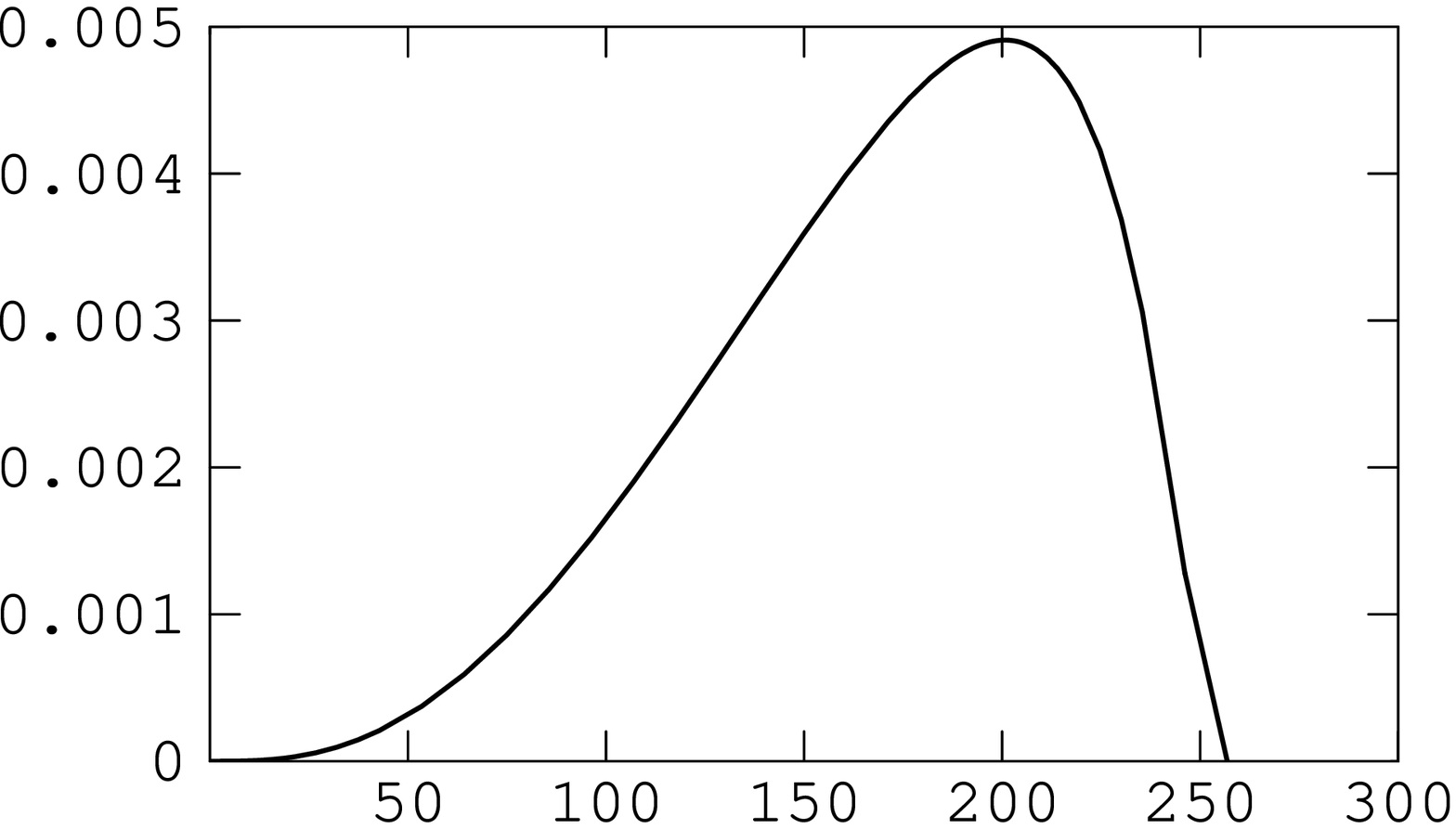}
\begin{flushright}
  \vspace*{-8mm}
   [MeV]
\end{flushright}
\caption{Spectrum of the photon energy $E_{\gamma}$.}
\end{figure}

Spectrum of the photon energy $E_{\gamma}$ for the \epg decay is shown in
Fig 4, and given in \cite{Ame} in ChPT.
Both are also similar, though there is no experimental result.

In our calculation of the \epg decay, we evaluate only the quark-box diagram
in Fig 1. Since the vector and axialvector four-quark interactions are not 
included in our model, the only other contribution to this process is the
scalar resonance exchange. In the ENJL model the contribution of the scalar
resonance exchange is small\cite{Bel}. We expect that similar result will be
obtained in our approach.

If one includes the vector and axialvector four-quark interaction in the NJL
model, the pseudoscalar meson properties are affected through the 
pseudoscalar-axialvector channel mixing and the model parameters with and
without the vector and axialvector four-quark interaction are different.
We expect that the models with and without the vector-axialvector interaction
predict similar results for the processes involving only the pseudoscalar
mesons with energies much below the vector meson masses. It is further
argued that the contribution of the quark-box diagram to the $\gamma \gamma
\to \pi^0 \pi^0$ process, that is similar to \epg, is quite close to that of
the vector meson exchange in the vector dominance model\cite{Bijnens2}.

In summary, we have studied the \epg  decay in the three-flavor NJL model
that includes the $U_{A}(1)$ breaking six-quark determinant interaction.
The $\eta$ meson mass, the $\eta \to \gamma \gamma$ decay width and the \epg
decay width are reproduced well with a rather strong $U_{A}(1)$ breaking
interaction, that makes $\eta_{1}-\eta_{8}$ mixing angle $\theta 
\simeq 0^{\circ}$. Since the $\eta'$ meson is expected to be sensitive to the
effects of the $U_{A}(1)$ anomaly, it is very important to study the $\eta'$
meson properties. It should be noted, however, that the NJL model does not
confine quarks. While the NG bosons, $\pi, K$ and $\eta$, are strongly
bound and therefore can be described in the NJL model fairly well, 
we do not apply our model
to the heavy mesons such as $\rho, \omega$ and $\eta'$. 
Further study of the $U_{A}(1)$ breaking and the $\eta_{1}-\eta_{8}$ mixing
will require a calculation including the confinement mechanism.


\begin{thebibliography}{99}
\bibitem{Taki}
  M. Takizawa and M. Oka, Phys. Lett. {\bf B 359} (1995) 210; {\bf B 364} 
(1995) 249 (E).

\bibitem{Hatsuda}
  T. Hatsuda and T. Kunihiro, Phys. Rep. {\bf 247} (1994) 221.

\bibitem{Don}
  J.F. Donoghue, B.R. Holstein, and Y.-C.R. Lin, Phys. Rev. Lett {\bf 55} 
(1985) 2766;  {\bf 61} (1988) 1527 (E).

\bibitem{Exp}
  Particle Data Group, M. Aguilar-Benitez et al, Phys. Rev. {\bf D 50} (1994) 
1173.

\bibitem{Ng}
  J.N. Ng and D.J. Peters, Phys. Rev. {\bf D 47} (1993) 4939.

\bibitem{Eck}
  G. Ecker, A. Pich and E. de Rafael, Nucl. Phys. {\bf B 303} (1988) 665.

\bibitem{Ame}
  Ll. Ametller, J. Bijnens, A. Bramon and F. Cornet, Phys. Lett. {\bf B 276} 
(1992) 185.

\bibitem{Ko2}
  P. Ko, Phys. Rev. {\bf D 47} (1993) 3933.

\bibitem{Ko}
  P. Ko, Phys. Lett. {\bf B 349} (1995) 555.

\bibitem{Bel}
  S. Bellucci and C. Bruno, Nucl. Phys. {\bf B 452} (1995) 626.

\bibitem{Bijnens}
  J. Bijnens, C. Bruno and E. de Rafael, Nucl. Phys. {\bf B 390} (1993) 501.

\bibitem{Bijnens2}
  J. Bijnens, S. Dawson and G. Valencia, Phys. Rev. {\bf D 44} (1991) 3555. 

\end{thebibliography}
\end{document}